\documentclass[pdftex,iop,apjl]{emulateapj} % Linewraps are better with apjl option

\usepackage{longtable}
\usepackage{epstopdf}  % allows you to use the pdflatex script from JCV
\usepackage{ifthen}
\usepackage{rotating}
\usepackage{subfigure} % allows numbering figures 1a, 1b, etc.
\usepackage{booktabs}

\usepackage{appendix}
\usepackage{color}

\submitted{Accepted for publication in The Astronomical Journal, 11 May 2015}
\shorttitle{{\sc Combined Investigation of Comet 103P/Hartley 2}}
\shortauthors{{\sc Knight et al.}}

\def\icarus{Icarus}

\begin{document}
\bibliographystyle{apj}

\title{A Further Investigation of Apparent Periodicities and the Rotational State of Comet 103P/Hartley 2 from Combined Coma Morphology and Lightcurve Datasets}

\author{Matthew M. Knight\altaffilmark{1,2,3}, {Beatrice E.A. Mueller\altaffilmark{4}}, {Nalin H. Samarasinha\altaffilmark{4}}, {David G. Schleicher\altaffilmark{2}}}

\altaffiltext{1}{Contacting author: knight@lowell.edu.}
\altaffiltext{2}{Lowell Observatory, 1400 W. Mars Hill Rd, Flagstaff, AZ 86001, USA}
\altaffiltext{3}{Visiting scientist at The Johns Hopkins University Applied Physics Laboratory, 11100 Johns Hopkins Road, Laurel, MD 20723, USA}
\altaffiltext{4}{Planetary Science Institute, 1700 E. Ft. Lowell, \#106, Tucson, AZ 85719, USA}

\begin{abstract}
We present an analysis of Kitt Peak National Observatory and Lowell Observatory observations of comet 103P/Hartley 2 obtained from August through December 2010. The results are then compared with contemporaneous observations made by the {\it EPOXI} spacecraft. Each ground-based dataset has previously been investigated individually; the combined dataset has complementary coverage that reduces the time between observing runs and allows us to determine additional apparent periods at intermediate times. We compare CN coma morphology between ground-based datasets, making nine new measurements of apparent periods. The first five are consistent with the roughly linearly increasing apparent period during the apparition found by previous authors. The final four suggest that the change in apparent period slowed or stopped by late November. We also measure an inner coma lightcurve in both CN and R-band ground-based images, finding a single-peaked lightcurve which repeats in phase with the coma morphology. The apparent period from the lightcurve had significantly larger uncertainties than from the coma morphology, but varied over the apparition in a similar manner. Our ground-based lightcurve aligns with the published {\it EPOXI} lightcurve, indicating that the lightcurve represents changing activity rather than viewing geometry of structures in the coma. The {\it EPOXI} lightcurve can best be phased by a triple-peaked period near 54--55 hr that increases from October to November. This phasing reveals that the spacing between maxima is not constant, and that the overall lightcurve shape evolves from one triple-peaked cycle to the next. These behaviors suggest that much of the scatter in apparent periods derived from ground-based datasets acquired at similar epochs are likely due to limited sampling of the data.

\end{abstract}

\keywords{comets: general --- comets: individual (103P/Hartley 2) --- methods: data analysis --- methods: observational}

\section{INTRODUCTION}
Comet 103P/Hartley 2 was extensively studied during the second half of 2010 due to it being the target of the {\it EPOXI} flyby on 2010 November 4 as well as having a close approach to Earth with favorable viewing geometry. {\it EPOXI} spacecraft observations revealed that the nucleus of Hartley 2 was much smaller than previously thought, with its apparent  ``hyperactivity'' due to a population of icy grains in the coma (e.g., \citealt{ahearn11a,hermalyn13,kelley13,protopapa14}). Details of the nucleus's shape and surface properties were determined (e.g., \citealt{groussin13,li13c,thomas13}) and used to explain the origins of its jet activity (e.g., \citealt{belton13b,brucksyal13}). Earth-based observations over multiple orbits revealed a secular decrease in activity since 1991 and a strong asymmetry in production rates around perihelion \citep{combi11b,knight13}. Many larger-scale properties at or near the time of the flyby were constrained from Earth, including an Earth ocean-like D/H ratio \citep{hartogh11}, activity levels and relative abundances of various gas species (e.g., \citealt{dellorusso11,mumma11,weaver11,crovisier13,mckay13,boissier14}), and coma morpology (e.g., \citealt{knight11b,lara11a,samarasinha11,waniak12,mueller13}). A number of measurements of the apparent period were determined and are discussed in detail below.

We obtained independent observations of Hartley 2 at Kitt Peak National Observatory (KPNO) and Lowell Observatory on complementary epochs from August 2010 through January 2011 which have already yielded much novel information about Hartley 2. Data from each observatory were analyzed separately, with repetition of CN coma features used to determine apparent periods \citep{iauc9163a,iauc9178}. As the apparition progressed, continued monitoring of the CN coma morphology revealed that the apparent period was increasing and suggested that the nucleus was in non-principal axis (NPA) rotation \citep{iauc9178,samarasinha11,knight11b} with coma morphology best matching multiples of three cycles apart \citep{knight11b}. We later studied the dust morphology, identifying two faint sunward-facing dust jets that behaved very differently from the strong CN gas jets and suggesting that they originated from different source regions on the nucleus \citep{mueller13}. We analyzed the coma morphology of other gas species, finding that the distributions of C$_2$ and C$_3$ were similar to that of CN, but the distributions of OH and NH were different and were likely derived from small icy grains that were subject to radiation pressure \citep{knight13}.

Numerous observers \citep{ahearn11a,cbet2589,knight11b,meech11,samarasinha11} showed that the apparent period near 17 hr in late August/early September slowed by an unusually large amount in only a few months. This change cannot always have been linear since the apparent period was nearly unchanged from mid-2009 \citep{meech11} until August 2010 \citep{iauc9163a}; this is unsurprising since period changes would be expected to be small or nonexistant due to lack of torques when activity was weak at large heliocentric distances. \citet{ahearn11a} and \citet{belton13a} identified several periodicities in the {\it EPOXI} data ranging from 17--90 hr. In addition to a periodicity corresponding to the lengthening $\sim$17 hr periodicity just discussed, they found a strong periodicity near 55 hr (or possibly $\sim$27 hr). This $\sim$55 hr periodicity remained constant or increased slightly from mid-October until late November, but its behavior in September was inconclusive. No evidence for or against a change in the longer apparent period has been detected via coma morphology. There is no published evidence for the extent of any change in the long apparent periods after the {\it EPOXI} observing interval. Understanding the behaviors of the short ``single-peaked'' and long ``triple-peaked'' apparent periods over as wide a range of dates as possible is critical for properly interpreting observations of Hartley 2 since it may help to reveal the underlying component periods of NPA rotation\footnote{Note that there are three component periods of NPA rotation: P$_\psi$, P$_\theta$, P$_\phi$, that P$_\psi$ and P$_\theta$ are coupled to each other, and that we are using the ``L-convention'' for defining the Euler angles (\citealt{samarasinha15} and references therein).}.

Despite the wealth of studies on Hartley 2, many uncertainties remain, and continued mining of existing datasets is worth pursuing. Hartley 2 remains a prime target for future spacecraft missions and many of the behaviors it exhibited were also seen in another prominent spacecraft target, 1P/Halley (cf. \citealt{millis86,schleicher90}). It has been well established that Hartley 2 is in NPA rotation (e.g., \citealt{ahearn11a}), but the mode of rotation, long-axis mode (LAM) versus short-axis mode (SAM), is not yet definitively known. \citet{belton13a} argued for a LAM while \citet{samarasinha12} pointed out that a SAM cannot be excluded. While determining the specifics of the NPA rotation are beyond the scope of this paper, our ultimate goal is to derive the specific rotational state and component periods of NPA rotation. Therefore, in this paper we are investigating apparent periodicities in the coma morphology (by observing the motion and variation of morphological features and looking for repetitions of patterns) and in the lightcurve (by measuring brightness variations of the inner coma) and how they evolve with time in order to constrain the rotational state. The apparent repeatability of morphology and/or lightcurve features is not necessarily a direct representation of the ``rotation'' of the nucleus, but may be caused by the orientation of the source region(s) on the nucleus appearing nearly the same at two separate intervals of time due to a combination of NPA rotation, changes in the component periods of NPA rotation, changing viewing geometry, and seasonal changes in activity. Thus, we avoid using the term ``rotation period'' when referring to apparent periods.

In this paper we combine KPNO and Lowell imaging data to gain new insight into the rotation of Hartley 2. The datasets were temporally offset from one another except at the time of the {\it EPOXI} encounter, and the combined dataset allows us to determine apparent periods from morphology at intermediate times that are impossible to derive for either individual dataset. The combined dataset provides additional lightcurve data at times beyond the {\it EPOXI} lightcurve coverage of Hartley 2 and which have not been reported in the literature by any other observers. Previous studies have yielded systematic offsets between apparent periods determined from lightcurve and coma morphology, e.g., between early September and early November, periods from coma morphology \citep{cbet2589,knight11b,samarasinha11} are higher than those from lightcurves at similar epochs \citep{ahearn11a,drahus11,waniak12,belton13a}. Our combined dataset allows the determination of apparent periods for lightcurve and coma morphology from the same data to test for differences. We also utilize the {\it EPOXI} lightcurve data to better understand our data and explore possible biases in apparent periods determined from more sparsely sampled Earth-based datasets. 

In Section~\ref{sec:observations} we briefly summarize the observations and reductions of each dataset. In Sections~\ref{sec:morphology} and \ref{sec:lightcurves} we present studies of the coma morphology and inner coma lightcurves, respectively. Finally, we discuss the implications of these new results in Section~\ref{sec:disc}.

\section{OBSERVATIONS AND REDUCTIONS}
\label{sec:observations}

As described previously in \citet{samarasinha11} and \citet{mueller13}, observations with the KPNO 2.1-m telescope had a field of view of 6.5\,arcmin $\times$ 10\,arcmin and a binned pixel scale of 0.305\,arcsec. Lowell observations were acquired with the Hall 42-in (1.1 m) and 31-in (0.8 m) telescopes, which had square fields of view of 25.3 arcmin and 15.7 arcmin, respectively, and binned pixel scales of 1.48 arcsec and 0.92 arcsec, respectively \citep{knight11b,knight13}. All images were trailed at the comet's rate of motion. Broadband R and HB narrowband comet filters \citep{farnham00} were used on all nights, however, we utilize only broadband R or narrowband CN images throughout this paper. Images in additional filters were acquired but generally had worse signal to noise than R or CN and were taken with insufficient frequency for the studies herein. All 31-in data were snapshots and were, therefore, only used for coma morphology assessment. Images were processed using standard bias subtraction and flat fielding techniques. The observing circumstances and weather conditions are given in Table~\ref{t:imaging_circ}.

\renewcommand{\baselinestretch}{0.6}
\renewcommand{\arraystretch}{1.5}

% TABLE 1
% Table of geometric Geometric circumstances during imaging
\begin{deluxetable*}{lccrcccccc}
\tabletypesize{\scriptsize}
\tablecolumns{12}
\tablewidth{0pt} 
\setlength{\tabcolsep}{0.05in}
\tablecaption{Summary of Hartley 2 imaging observations and geometric parameters in 2010.\,\tablenotemark{a}}
\tablehead{   % column headings
  \colhead{UT Date}&
  \colhead{UT Range}&
  \colhead{Tel\tablenotemark{b}}&
  \colhead{$\Delta$T\tablenotemark{c}}&
  \colhead{$r_\mathrm{H}$\tablenotemark{d}}&
  \colhead{$\Delta$\tablenotemark{e}}&
  \colhead{$\theta$\tablenotemark{f}}&
  \colhead{PA Sun\tablenotemark{g}}&
  \colhead{Conditions}&
  \colhead{Used in}\\
  \colhead{}&
  \colhead{}&
  \colhead{}&
  \colhead{(days)}&
  \colhead{(AU)}&
  \colhead{(AU)}&
  \colhead{($^\circ$)}&
  \colhead{($^\circ$)}&
  \colhead{}&
  \colhead{Lightcurves?}
}
\startdata
Aug 13&03:28--11:48&42in&$-$75.9&1.463&0.567&30&22&Photometric&Yes\\
Aug 14&03:03--11:48&42in&$-$74.9&1.454&0.557&30&21&Photometric&Yes\\
Aug 15&03:15--12:10&42in&$-$73.9&1.446&0.547&30&20&Clouds&Yes\\
Aug 16&07:40--11:54&42in&$-$72.8&1.437&0.536&31&19&Clouds&Yes\\
Aug 17&07:48--11:01&42in&$-$71.9&1.421&0.527&31&18&Clouds&Yes\\
Sep 1&03:24--11:47&KP2.1&$-$56.9&1.310&0.391&34&3&Photometric&Yes\\
Sep 2&03:22--12:01&KP2.1&$-$55.9&1.302&0.383&34&3&Cirrus&Yes\\
Sep 3&03:08--11:04&KP2.1&$-$55.0&1.295&0.375&35&2&Cirrus&Yes\\
Sep 9&02:53--12:16&42in&$-$48.9&1.252&0.328&37&357&Photometric&Yes\\
Sep 10&02:36--12:09&42in&$-$48.0&1.245&0.320&37&357&Photometric&Yes\\
Sep 11&02:33--12:08&42in&$-$47.0&1.238&0.313&37&356&Photometric&Yes\\
Sep 12&02:33--12:11&42in&$-$46.0&1.231&0.305&38&356&Clouds&Yes\\
Sep 13&02:35--12:11&42in&$-$44.9&1.224&0.298&38&355&Clouds&Yes\\
Sep 30&03:19--12:12&KP2.1&$-$27.9&1.127&0.188&44&3&Patchy clouds&Yes\\
Oct 1&02:25--11:30&KP2.1&$-$27.0&1.123&0.183&45&5&Patchy clouds&Yes\\
Oct 2&03:53--12:24&KP2.1&$-$25.9&1.118&0.177&45&7&Patchy clouds&Yes\\
Oct 3&05:43--12:36&KP2.1&$-$24.9&1.114&0.172&46&10&Cloudy&Yes\\
Oct 4&03:06--09:02&KP2.1&$-$24.0&1.110&0.167&46&12&Cirrus&Yes\\
Oct 12&03:12--12:34&31in&$-$15.9&1.082&0.134&49&43&Photometric&No\\
Oct 13&03:13--12:43&31in&$-$14.9&1.079&0.131&49&48&Photometric&No\\
Oct 14&03:13--12:40&31in&$-$13.9&1.076&0.129&50&52&Photometric&No\\
Oct 15&03:06--05:33&31in&$-$13.1&1.074&0.127&50&56&Clouds&No\\
Oct 16&05:01--12:21&42in&$-$11.9&1.072&0.125&51&61&Thin cirrus&Yes\\
Oct 17&05:00--12:38&42in&$-$10.9&1.070&0.123&51&65&Clouds&Yes\\
Oct 19&10:56--12:24&42in&$-$8.8&1.066&0.121&52&73&Clouds&No\\
Oct 31&07:10--12:36&31in&$+$3.2&1.060&0.140&58&99&Thin cirrus&No\\
Nov 1&07:15--12:45&31in&$+$4.2&1.060&0.144&59&100&Thin cirrus&No\\
Nov 2&07:37--12:35&KP2.1&$+$5.2&1.061&0.148&59&102&Photometric&Yes\\
Nov 2&06:45--12:54&42in&$+$5.2&1.061&0.147&59&102&Photometric&Yes\\
Nov 2&07:27--10:32&31in&$+$5.1&1.061&0.147&59&102&Photometric&No\\
Nov 3&07:47--12:53&KP2.1&$+$6.2&1.062&0.152&59&103&Cirrus&Yes\\
Nov 3&06:41--13:01&42in&$+$6.2&1.062&0.151&59&103&Photometric&Yes\\
Nov 4&07:48--12:49&KP2.1&$+$7.2&1.063&0.156&59&104&Cirrus&Yes\\
Nov 4&06:39--13:07&42in&$+$7.2&1.063&0.155&59&104&Thin cirrus&Yes\\
Nov 5&07:50-12:54&KP2.1&$+$8.2&1.065&0.160&59&105&Cirrus&Yes\\
Nov 5&07:44--10:45&31in&$+$8.1&1.065&0.159&59&105&Photometric&No\\
Nov 6&07:50--12:54&KP2.1&$+$9.2&1.066&0.164&59&106&Cirrus&Yes\\
Nov 6&07:41--10:58&31in&$+$9.1&1.066&0.163&59&106&Thin cirrus&No\\
Nov 7&07:45--12:53&KP2.1&$+$10.2&1.068&0.168&59&107&Patchy clouds&Yes\\
Nov 7&06:48--13:09&42in&$+$10.2&1.068&0.168&59&107&Photometric&Yes\\
Nov 8&08:04--12:46&KP2.1&$+$11.2&1.070&0.173&59&108&Photometric&Yes\\
Nov 10&07:59--13:11&31in&$+$13.2&1.074&0.181&58&109&Clouds&No\\
Nov 12&07:56--12:30&31in&$+$15.2&1.079&0.191&58&111&Clouds&No\\
Nov 13&08:09--13:10&31in&$+$16.2&1.082&0.196&57&112&Clouds&No\\
Nov 16&08:08--13:08&31in&$+$19.2&1.091&0.211&56&114&Photometric&No\\
Nov 26&07:45--12:43&31in&$+$29.2&1.132&0.262&51&123&Photometric&No\\
Nov 27&07:48--12:46&31in&$+$30.2&1.137&0.267&50&124&Photometric&No\\
Dec 9&07:02--13:19&42in&$+$42.2&1.205&0.330&42&136&Thin cirrus&Yes\\
Dec 10&06:48--08:50&42in&$+$43.1&1.212&0.336&41&137&Thin cirrus&Yes\\
Dec 11&07:23--12:11&KP2.1&$+$44.1&1.218&0.342&41&138&Cloudy&Yes\\
Dec 12&07:11--11:59&KP2.1&$+$45.1&1.225&0.347&40&139&Cirrus&Yes\\
Dec 13&07:17--12:02&KP2.1&$+$46.1&1.231&0.353&39&141&Cirrus&Yes\\
Dec 14&06:54--11:58&KP2.1&$+$46.9&1.238&0.358&39&142&Photometric&Yes\\
Dec 15&07:00--11:58&KP2.1&$+$48.1&1.245&0.364&38&143&Photometric&Yes\\
Dec 15&08:57--09:13&42in&$+$48.1&1.245&0.363&38&143&Clouds&No\\
\vspace{-1mm}
\enddata
\tablenotetext{a} {All parameters are given for the midpoint of each night's observations.}
\tablenotetext{b} {KP2.1 = KPNO 2.1-m Telescope, 42in = Lowell Observatory Hall 42-in Telescope (1.1 m), 31in = Lowell Observatory 31-in Telescope (0.8 m).}
\tablenotetext{c} {Time relative to perihelion.}
\tablenotetext{d} {Heliocentric distance.}
\tablenotetext{e} {Geocentric distance.}
\tablenotetext{f} {Solar phase angle.}
\tablenotetext{g} {Position angle of the Sun, measured from north through east (counterclockwise when displayed with north up and east to the left).}
\label{t:imaging_circ}
\end{deluxetable*}

\renewcommand{\baselinestretch}{0.9}
\renewcommand{\arraystretch}{1.0}

For morphological analysis, we employed enhancement techniques to remove the bulk background and make the fainter features more obvious. KPNO images were enhanced by dividing by each image's azimuthal average profile (with high pixel rejection) and Lowell images were enhanced by dividing by each image's azimuthal median profile. These enhancement techniques produce nearly identical results, a fact which was confirmed by comparison of enhanced images for all nights on which simultaneous data were obtained. We created subimages of the enhanced images that were centered on the nucleus and had the same physical scale throughout the apparition, $\sim$50,000 km at the comet. Because of the smaller field of view of the KPNO data, the outer portion of this region was not always observed and these regions were set to 0. 

Aperture photometry was conducted using two apertures that were fixed in physical size at the comet (radii of 1000 and 2000 km) and five apertures that were fixed in angular scale (radii of 3, 6, 9, 12, and 15 arcsec). Larger apertures had more signal but also diluted the intrinsic variability and suffered more contamination from background stars. Conversely, smaller apertures reduced stellar contamination and showed larger amplitude variations, but had less signal and, if too small, did not necessarily encompass all of the feature(s) causing the brightness variation. Ultimately the 9 arcsec radius aperture proved to be the optimum size and was used for the lightcurve analysis. Appropriately large background annuli were utilized for sky subtraction for each run, but there was likely always some coma present in the background annulus. Coma contamination diminishes the lightcurve amplitude and affects night to night normalization, but does not affect the timing of extrema and therefore was of minimal concern. 

We did relative photometry with on-chip stars as outlined in \citet{knight11a} for all Lowell images and for KPNO images from December, allowing us to recover a useful lightcurve on non-photometric nights for extinction $<$0.5 mag. Due to the combination of a small field of view and the comet's high rate of motion, this was not possible for KPNO images from September, October, and November. Therefore non-photometric nights during these months could not be used for R-band lightcurve analysis since the uncertainties due to clouds were of order the R-band amplitude. We have included CN lightcurves from these nights because they have significantly larger amplitudes, but we were cautious of over-interpreting unusual lightcurve behavior on these nights.

\section{CN COMA MORPHOLOGY}
\label{sec:morphology}
As discussed in \citet{samarasinha11} and \citet{knight11b}, large scale features were evident in the CN coma that changed smoothly during a night and varied from night to night. The repetition of these features on different nights within a given run previously allowed us to determine the apparent period. The CN images were not significantly contaminated by either continuum or other gas species, allowing us to use data from non-photometric nights and greatly increasing the number of nights available for morphological comparison than if we only used photometric nights. The only large scale feature seen in the R images was the dust tail which varied little from night to night. Two relatively short, sunward facing features were also seen in the highest signal to noise images (e.g., \citealt{mueller13}). While their orientations changed over diurnal time scales, these variations were insufficient to constrain the apparent period from our two datasets (Samarasinha et al., in prep., are attempting to do this using observations from locations around the world; see also \citealt{samarasinha12}) and so R-band morphology is not discussed further.

\begin{figure}[ht]
  \centering
  \includegraphics[width=88mm]{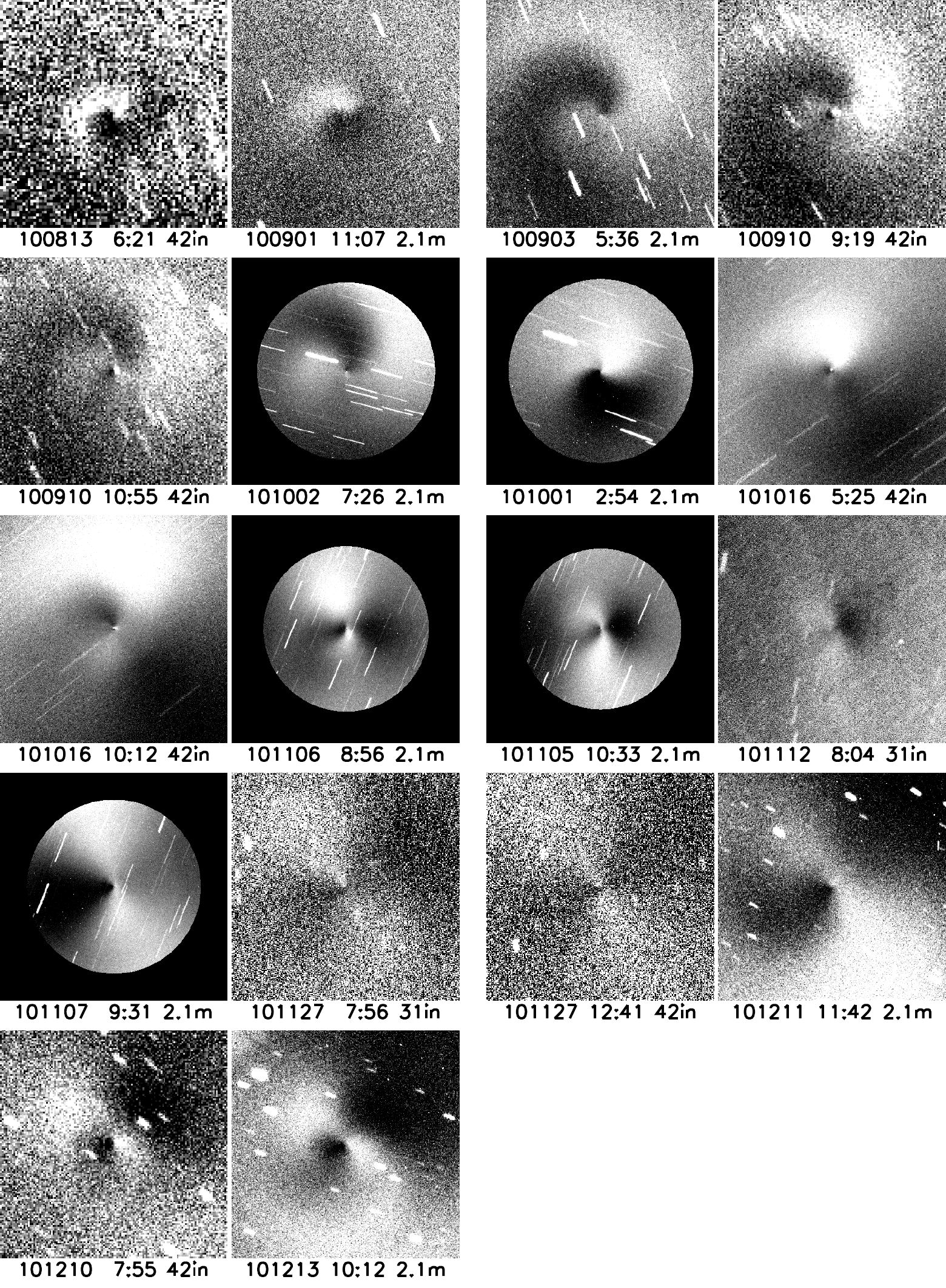}
  \caption[Matched pairs of images]{\footnotesize{Pairs of matching CN images from different observing runs. The date (YYMMDD), UT time (HH:MM), and telescope are given below each image. The time sequence moves from the left to right pair and then from top to bottom. Each image is centered on the nucleus, has been enhanced as described in the text, has north up and east to the left, and is $\sim$50,000 km across at the comet. The grayscale stretch is the same for both images in a pair, but is different from pair to pair. White is bright and black is dark, and trailed stars appear as diagonal streaks in some images. Some 2.1-m images have a circular black border; these are regions where no data were obtained due to the smaller field of view. Slight mismatches in orientation are primarily due to geometry changes between runs, e.g., between October 16 and November 6.}}
  \label{fig:cn_pairs}
\end{figure}

As with our previous investigations of the apparent period within an individual run, we compared the shape, size, radial distance, orientation, and relative intensity of CN coma features in order to find pairs of images with similar morphology. Here we concerned ourselves only with images in previously uncompared datasets, e.g. Lowell data from August 13--17 with KPNO data from September 1--3, KPNO data from September 1--3 with Lowell data from September 10--13, etc. We compared all images in consecutive datasets and noted all pairs of images with similar morphology. Selected pairs are shown in Figure~\ref{fig:cn_pairs}. Given the large baseline between observing runs and the reasonably high cadence, typically $\sim$20 min for KPNO and $\sim$60 min for Lowell, interpolation between images was not necessary.

%FIGURE 1: selected pairs of images

\begin{figure}[t]
  \centering
  \includegraphics[width=87mm]{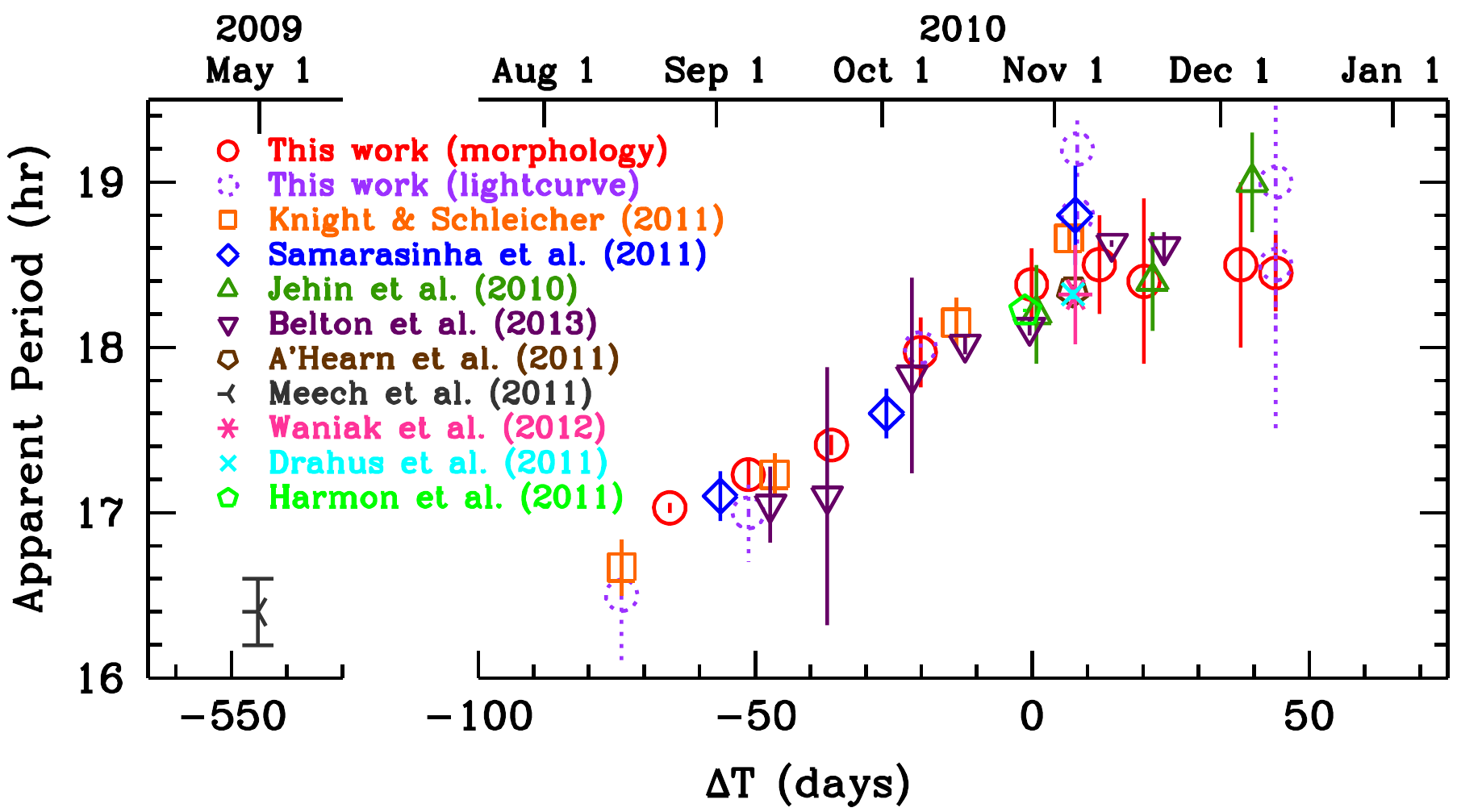}
  \caption[Increasing period]{Apparent periods of Hartley 2 as a function of time relative to perihelion ($\Delta$T). Symbols are defined in the legend. Coma morphology was used to determine apparent periods in this work (Table~\ref{t:period_meas}), \citet{cbet2589}, \citet{knight11b}, and \citet{samarasinha11}. Apparent periods were determined from photometric measurements by \citet{drahus11}, \citet{meech11}, \citet{ahearn11a}, \citet{waniak12}, \citet{belton13a}, and this work (Table~\ref{t:period_phased}; both CN and R results are plotted with the same symbols). \citet{harmon11} used radar observations of the nucleus. Error bars are shown but do not necessarily have the same significance between different data sets due to differing methodologies. Note that the upper error bar on the lightcurve point at ${\Delta}T=+$44 days is truncated and actually extends to 20.5 hr. 
}
  \label{fig:rot_period}
  \label{lastfig}
\end{figure}

We determined the apparent period between pairs of images by dividing the time between images by the integer number of cycles that yielded a period closest to the expected average period based on extrapolation from previously published results. Solutions of $N+$1 cycles and $N-$1 cycles were also considered for an initial solution of $N$ cycles. However, in most cases the $N\pm$1 cycle solutions had larger standard deviations, yielded periods that were beyond the error bars of previously published solutions, and were inconsistent with a smoothly changing period. These were therefore considered  unlikely. Furthermore, sometimes the $N\pm$1 solutions could be ruled out within a single run, e.g., when data were obtained 2--3 days apart. The final solution for the period between two observing runs was the mean period of all measured pairs between the two runs and the 1-$\sigma$ uncertainty was the standard deviation of the measured periods. The time at which this mean period applied was the mean of the midpoints of all pairs. We present a summary of our period solutions in Table~\ref{t:period_meas}, including the $N+$1 and $N-$1 solutions, and plot our preferred solutions along with other published periods in Figure~\ref{fig:rot_period}. The periods centered on November 9.40, November 17.45, and December 4.93 were calculated from three or fewer pairs of images. While the formal uncertainty in these periods is less than 0.1 hr, we consider the real error to be no better than 0.5 hr, and round to the neareast 0.1 hr.

% Table 2 - period from morphology
\begin{deluxetable*}{lllccccc}  
\tabletypesize{\scriptsize}
\tablecolumns{5}
\tablewidth{0pt} 
\setlength{\tabcolsep}{0.05in}
\tablecaption{Apparent periods measured from coma morphology}
\tablehead{   % column headings
  \colhead{Date}&
  \colhead{Date}&
  \colhead{Average}&
  \colhead{Intervening}&
  \colhead{Num.}&
  \colhead{Period (hr)}&
  \colhead{Period (hr)}&
  \colhead{Period (hr)}\\
  \colhead{Range \#1}&
  \colhead{Range \#2}&
  \colhead{Date}&
  \colhead{Cycles}&
  \colhead{Pairs}&
  \colhead{$N$ cycles}&
  \colhead{$N+$1 cycles}&
  \colhead{$N-$1 cycles}
}
\startdata
Aug 13--17&Sep 1--3&Aug 23.87&22--27&20&17.03 $\pm$ 0.03&16.41 $\pm$ 0.03&17.70 $\pm$ 0.08\\
Sep 1--3&Sep 10--13&Sep 7.04&10--17&70&17.23 $\pm$ 0.08&15.97 $\pm$ 0.19&18.71 $\pm$ 0.26\\
Sep 10--13&Sep 30--Oct 4&Sep 22.04&23--32&36&17.41 $\pm$ 0.06&16.82 $\pm$ 0.06&18.05 $\pm$ 0.10\\
Sep 30--Oct 4&Oct 12--16&Oct 8.13&12--20&16&17.97 $\pm$ 0.21&16.90 $\pm$ 0.35&19.20 $\pm$ 0.16\\
Oct 16--19&Nov 6--8&Oct 28.17&26--29&14&18.38 $\pm$ 0.22&17.74 $\pm$ 0.19&19.08 $\pm$ 0.25\\
Nov 3--6&Nov 10--16&Nov 9.40&7--14&3\tablenotemark{a}&18.5\phantom{0} $\pm$ 0.5\phantom{0}&16.6\phantom{0} $\pm$ 0.5\phantom{0}&20.8\phantom{0} $\pm$ 0.5\phantom{0}\\
Nov 7&Nov 27&Nov 17.45&26&2\tablenotemark{a}&18.4\phantom{0} $\pm$ 0.5\phantom{0}&17.8\phantom{0} $\pm$ 0.5\phantom{0}&19.2\phantom{0} $\pm$ 0.5\phantom{0}\\
Nov 27&Dec 11--13&Dec 4.93&18--21&2\tablenotemark{a}&18.5\phantom{0} $\pm$ 0.5\phantom{0}&17.6\phantom{0} $\pm$ 0.5\phantom{0}&19.5\phantom{0} $\pm$ 0.5\phantom{0}\\
Dec 9--10&Dec 10--13&Dec 11.24&1--5&6&18.45 $\pm$ 0.23&14.06 $\pm$ 2.45&24.05 $\pm$ 0.90
\enddata
\tablenotetext{a} {Periods given with 0.1 hr accuracy and uncertainty set to 0.5 hr rather than the formal uncertainty ($<$0.1) to avoid implying misleading precision due to small numbers of matched pairs of images.}
\label{t:period_meas}
\label{lasttable}
\end{deluxetable*}

%FIGURE 2: all published periods

It was more difficult to match features in the coma between observing runs than within individual observing runs because the longer baseline resulted in changes in the ratios of component periods, larger differences in viewing geometry,  etc. The coma morphology from mid-August through mid-September was reasonably similar. This allowed a large number of matching pairs of images with a high degree of confidence in the matches, resulting in a small standard deviation. However, the coma morphology from late September onward (perigee was on October 20) varied substantially and resulted in far fewer matching features from run to run and significantly less certainty about the mean period. Furthermore, as the coma features changed from a nearly face-on spiral in August to a side-on hour-glass shape by November, increasingly more subsets of images had little to no easily distinguishable coma features. The absence of distinct coma features meant that such images could not be used to determine matching pairs, although they were useful for eliminating alternate apparent periods.

As seen in Figure~\ref{fig:rot_period}, our newly measured apparent periods agree well with those previously published. The scatter in our periods compared to the ensemble is smaller than it was for either of our individual datasets, presumably due to the longer baseline between image pairs. The one exception is that the period determined for late August appears somewhat higher than expected relative to the individual periods in mid-August and early September, but is likely more robust because of the longer baseline. These new measurements are also more consistent with previous photometry-based periods, suggesting that there is not a systematic difference between techniques. The post-{\it EPOXI} flyby measurements have large uncertainties but, due to the number of epochs, give a stronger indication of a likely flattening of the rotation period than any previous results. We will return to these topics in more detail in Section~\ref{sec:disc}.

\section{LIGHTCURVES}
\label{sec:lightcurves}
Since Hartley 2 has a small nucleus and was highly active throughout our observations, its nucleus signal was swamped by the coma signal. The strong CN jets made us optimistic that there would be enough variation in the CN signal to measure a lightcurve. Furthermore, our observing cadences were high since we were looking for rotational variation in the CN coma morphology, so our CN temporal coverage was good. Since the dust is entrained in escaping gas, the potential large variation of the CN gas suggested that a corresponding variation in the R-band brightness was possible. The R-band images also had high signal to noise and good temporal coverage.

An inner coma lightcurve is typically much more complicated than a nucleus lightcurve, as it may depend on factors including the number and relative strength of the jets, the location of the jets, the duration of activity for each jet due to its local day/night, and projection effects (e.g., the lightcurve of 1P/Halley by \citealt{schleicher90}). Further complicating matters, NPA rotation means a jet will not necessarily be in the same orientation each ``rotational'' cycle. {\it EPOXI} obtained a coma lightcurve in the clear filter from September 5 through November 26 which exhibited a single-peaked sinusoid that varied in amplitude over three ``rotation'' cycles \citep{ahearn11a,belton13a}. Our data extend before and after the {\it EPOXI} data, but the {\it EPOXI} findings gave us reason to believe that the lightcurve during our observations would exhibit a similar behavior.

\begin{figure*}[t]
  \centering
  \includegraphics[width=180mm]{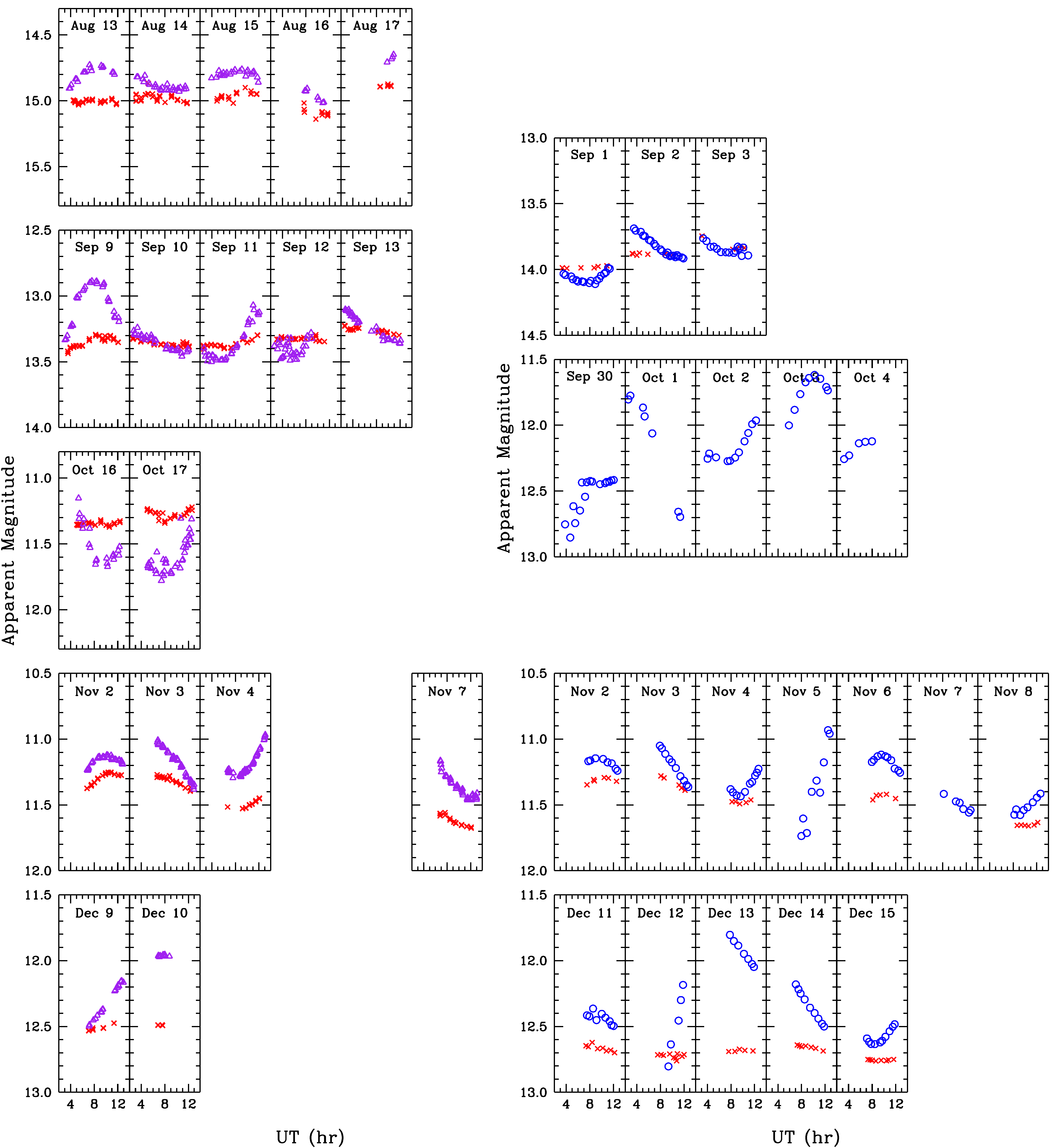}
  \caption[Nightly lightcurve by month]{\footnotesize{Nightly R and CN lightcurves by observing run for Lowell (left column) and KPNO (right column). R-band points are red crosses and CN points are purple triangles (Lowell) or blue circles (KPNO). All magnitudes are calculated in 9 arcsec radius apertures. Each point has been corrected for the light travel time. Magnitudes were corrected for extinction but not for geometric effects which vary significantly over time. Note that there were slight differences in the background removal between observatories due to the smaller chip size at KPNO. Error bars are not shown because they are smaller than the points; typical statistical uncertainties are $<$0.01 mag for R and $<$0.02 mag for CN. The magnitude range is different from panel to panel but the span is 1.5 dex in all panels. The temporal range is the same in all plots; note that there are gaps between nightly observations which are not shown. Therefore, data cannot be directly interpolated between the panels. Note that the lightcurve is unreliable on September 12 and the second half of September 13 due to clouds. KPNO R-band lightcurves are not shown for September 30--October 4, November 5, and November 7 because the nights were not photometric and the field of view was too small to use comparison stars; uncertainties due to cirrus would be on the order of the R-band amplitude but are significantly smaller than the observed CN amplitudes.
}}
  \label{fig:nightly_lc}
  \label{lastfig}
\end{figure*}

\begin{figure*}[t]
  \centering
  \includegraphics[width=180mm]{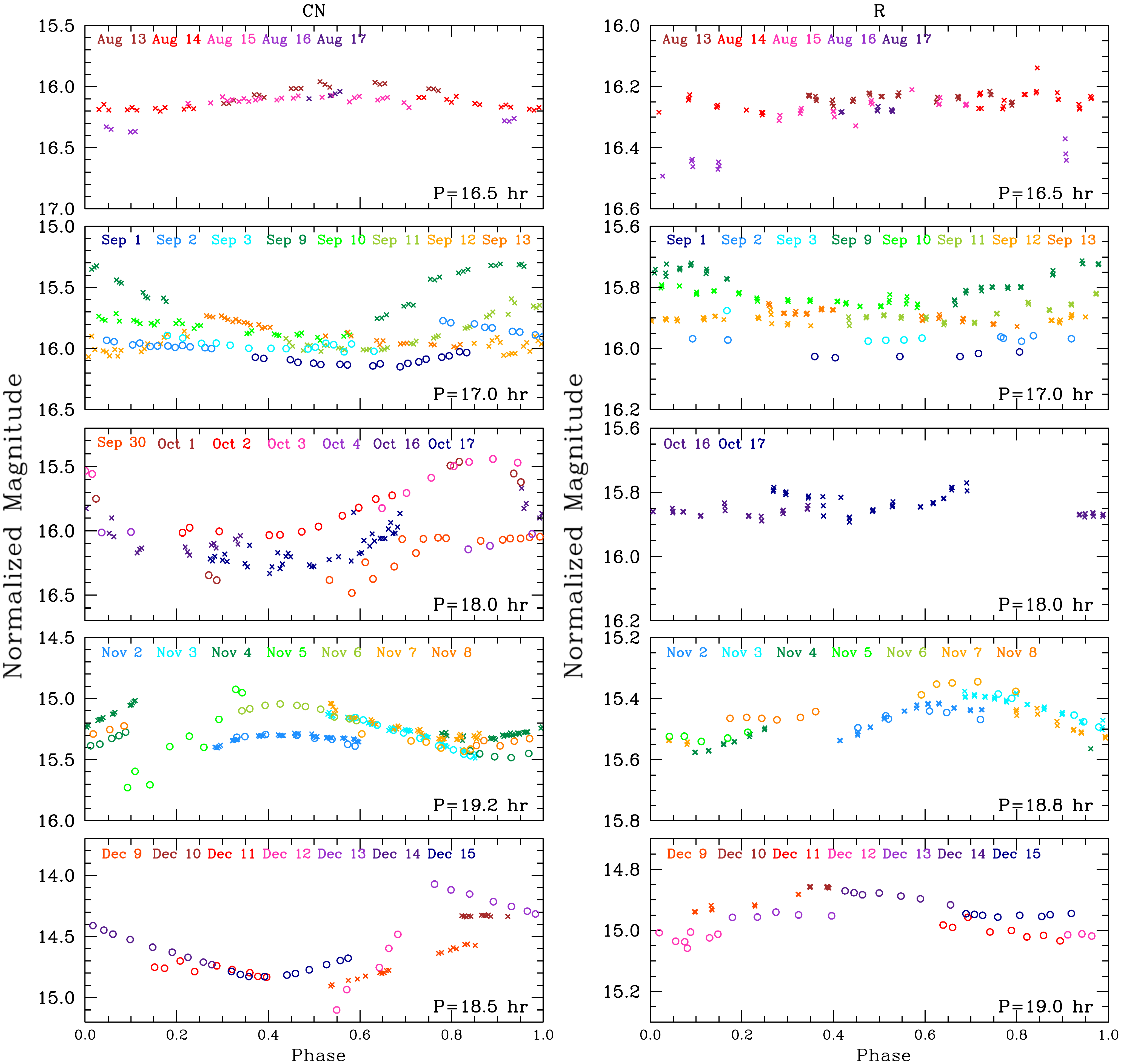}
  \caption[Phased lightcurve by month]{Rotationally phased lightcurves for each observing run or pair of runs. The left column is the CN lightcurve and the right column is the R lightcurve. The period to which the data are phased is given in the bottom right of each panel. Zero phase was set to perihelion in all plots. The nights plotted on a given panel are listed across the top of the panel and color coded to match the points. Each night has the same color on all panels in which it appears. Note that because the period by which the data are phased varies from panel to panel, the lightcurve features are not expected to align. Lowell data are plotted as crosses and KPNO data are plotted as open circles. All magnitudes have been normalized to ${\Delta}~=$~1~AU by an inverse square law. The magnitude range is different from panel to panel, but the span is 1.5 dex for all CN panels and 0.6 dex for all R panels.
}
  \label{fig:phased_lc}
  \label{lastfig}
\end{figure*}

The inner coma photometry is plotted in Figures~\ref{fig:nightly_lc} and \ref{fig:phased_lc}. Figure~\ref{fig:nightly_lc} shows the lightcurve as a function of UT on a night by night basis for each observing run, with R and CN plotted on the same panels. Figure~\ref{fig:phased_lc} shows the same data but phased to the best ``period'' for that run or pair of adjacent runs (the phasing will be discussed in Section~\ref{sec:lc_period}). For clarity, the CN and R lightcurves are plotted separately in Figure~\ref{fig:phased_lc}.

%FIGURE 3: Nightly lightcurves by month

%FIGURE 4: Phased lightcurves by month

\subsection{Lightcurve Shape and Amplitude}
As expected, the lightcurve displayed a single-peaked shape corresponding to fluctuations in brightness of the inner coma. The CN lightcurve displayed a much larger amplitude than the R lightcurve. It is difficult to measure the peak to trough variation precisely because no single night included both the lightcurve maximum and minimum and because the amplitude appears to vary from cycle to cycle. However, we estimate the peak to trough variation in a 9 arcsec aperture was at least 0.4--0.5 mag throughout the apparition for CN, while for R it was less than 0.1 mag in August and September and less than 0.2 mag in October, November, and December. These variations are smaller than those reported by \citet{ahearn11a} for grains, CO$_2$, and H$_2$O because the {\it EPOXI} lightcurves were measured in a much smaller aperture (191 km across) that responds to changes over very short timescales and results in larger amplitudes. 

The lower amplitude in R relative to CN is not surprising. Gas molecules (e.g., CN) are typically traveling at significantly higher velocities than are dust grains (e.g., the majority of the signal in the R images) and will therefore leave a given aperture more quickly. Furthermore, dust grains vary in size, and larger grains will travel slower than small grains, with the largest grains remaining near the nucleus for a long time. The higher velocity of gas and the dispersion of dust velocities combine to reduce the variability of the dust signal in the inner coma as compared to the gas.

\subsection{Apparent Period}
\label{sec:lc_period}
We determined the apparent period of our lightcurve data by plotting all data from a run, or a pair of runs if close together, phased to a particular period with zero phase set to the time of perihelion (2010 October 28.257), then changing the period slightly and redisplaying the phased lightcurve in real time. We looked for alignment of lightcurve features on different cycles but, owing to the varying amplitude and instrinsic brightness from night to night even when photometric, extrema needed only agree in phase to be considered aligned. We estimated the viable range of apparent periods as the times when the extrema first became noticeably out of phase. Visually phasing our CN lightcurves yielded periods at five epochs between mid-August and mid-December (some subsets of data used for coma periods were inconclusive for constraining the lightcurve period). Because there was much less variation in the R lightcurve, we were only able to constrain the period with it at two epochs, in November and December. The apparent CN and R lightcurve periods are given in Table~\ref{t:period_phased} and plotted in Figure~\ref{fig:rot_period}; the data are shown phased to our preferred apparent period for each run in Figure~\ref{fig:phased_lc}. We attempted to determine periods using Fourier analysis, but could not do so reliably due to the limitations of the dataset. Our phase coverage was insufficient to significantly constrain the longer (triple-peaked) period.

% Table 3 - period from lightcurves
\begin{deluxetable*}{lllcccc}  
\tabletypesize{\scriptsize}
\tablecolumns{7}
\tablewidth{0pt} 
\setlength{\tabcolsep}{0.05in}
\tablecaption{Apparent periods measured from phased lightcurves}
\tablehead{   % column headings
  \colhead{Date}&
  \colhead{Date}&
  \colhead{Midpoint}&
  \multicolumn{2}{c}{CN}&
  \multicolumn{2}{c}{R}\\
  \cmidrule(lr){4-5}
  \cmidrule(lr){6-7}
  \colhead{Range \#1}&
  \colhead{Range \#2}&
  \colhead{}&
  \colhead{Best (hr)}&
  \colhead{Range (hr)}&
  \colhead{Best (hr)}&
  \colhead{Range (hr)}
}
\startdata
Aug 13--17\tablenotemark{a}&...&Aug 15&16.5&16.1--16.7&...&...\\
Sep 1--3&Sep 10--13&Sep 7&17.0&16.9--17.2&...&...\\
Sep 30--Oct 4&Oct 16--17&Oct 8&18.0&18.0--18.1&...&...\\
Nov 2--8\tablenotemark{a}&...&Nov 5&19.2&19.0--19.4&18.8&18.6--18.9\\
Dec 9--15\tablenotemark{a}&...&Dec 12&18.5&18.1--18.6&19.0&17.5--20.5
\enddata
\tablenotetext{a} {Apparent periods determined from a single set of consecutive nights.}
\label{t:period_phased}
\label{lasttable}
\end{deluxetable*}

These periods are generally consistent with the periods we derived from the coma morphology (Section~\ref{sec:morphology}), but with larger uncertainties. The larger uncertainty is due to a combination of factors. First, we had numerous non-photometric nights which reduced the number of potential matches. These nights were generally salvageable by using comparison stars in the field of view. However, portions of nights with extinction greater than 0.5 mag and KPNO nights in October and November that had more than minimal cirrus (background stars moved quickly across the chip in these months) were effectively unusable for constraining the period. Also, some increased level of uncertainty was due to intermittently losing individual images because of contamination from background stars in the photometric aperture.  More problematic were the quirks of the lightcurve itself. The relatively simple single-peaked curve with a period near 16--19 hr meant that we could only sample a portion of the lightcurve on a given night, and the amount sampled decreased during the apparition as the nightly observing window shrunk towards the end of the year. If that section did not include a clear minimum or maximum it was very difficult to use to refine the period, although it was useful for excluding some periods. Furthermore, the lightcurve itself varied in amplitude and intrinsic brightness from one cycle to the next so sections without extrema could not be definitively compared. Thus, the lightcurve required many nights of data to be sampled sufficiently for phasing. Compounding this problem, the shape of lightcurve features was apparently not constant from one cycle to the next, as will be discussed in Section~\ref{sec:epoxi_lc_phased}. Finally, the time between extrema was apparently increasing throughout the apparition so it was difficult to use a longer baseline to improve the apparent period. We will discuss this last point in more detail below.

\subsection{Comparison with {\it EPOXI} Lightcurve}
\begin{figure}[t]
  \centering
  \includegraphics[width=87mm]{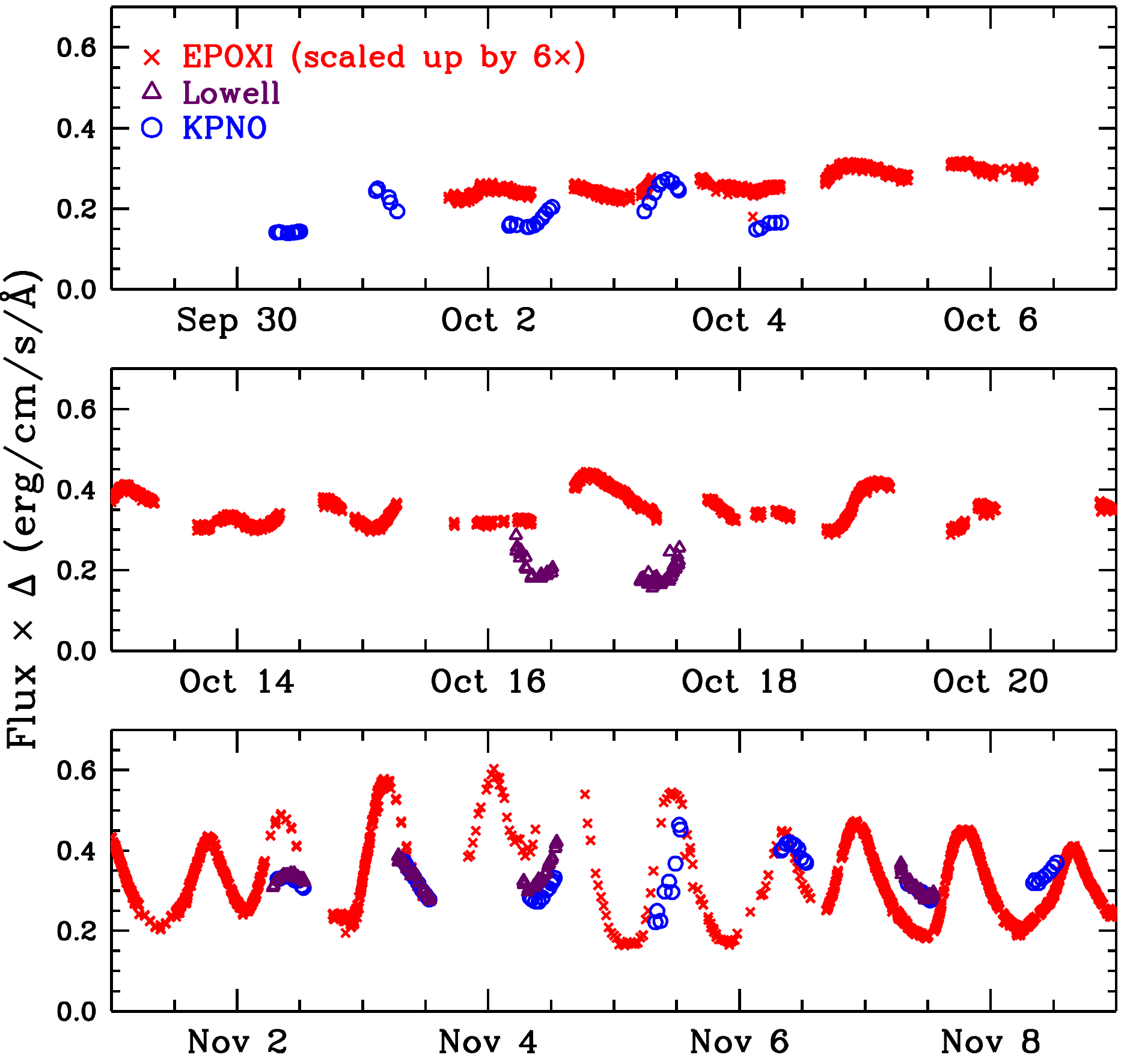}
  \caption[{\it EPOXI} $+$ ground-based lightcurves]{KPNO (blue circles) and Lowell (purple triangles) CN lightcurves overplotted with the {\it EPOXI} (red crosses) clear filter lightcurve from \citet{williams13}. {\it EPOXI} CN filter photometry is available but was acquired much less frequently than the clear filter so we have elected to display the clear filter data. All fluxes were extracted in apertures of radius 6 arcsec and were multiplied by the observer-centric distance ($\Delta$) to remove the effect of the changing geocentric distance. All times were corrected for light travel. The {\it EPOXI} data were scaled up by 6$\times$ to improve visibility; this had no effect on the timing of features in the lightcurve. The significantly smaller scale at the comet for {\it EPOXI} in November results in a much larger amplitude due to the decreased crossing time for grains in the aperture and the decrease in smearing out of the signal due to velocity disperson.}
  \label{fig:epoxi_lc}
\end{figure}

Calibrated aperture photometry from {\it EPOXI}'s Medium Resolution Visible CCD (MRI) images is available through the Planetary Data System for October 1--November 26, the interval when Hartley 2 was observed nearly continuously \citep{williams13}\footnote{While \citet{belton13a} analyzed the {\it EPOXI} lightcurve beginning on September 5, the earliest data available in the archive is from October 1.}. Thus, we were able to directly compare our inner coma lightcurve with that obtained by the spacecraft during our October and November runs. {\it EPOXI} MRI photometry is available in the OH, ultraviolet continuum, CN, C$_2$, green continuum, and clear filters (see \citealt{hampton05} for detailed filter information), but only the clear filter was sufficiently well sampled for direct comparison with our data. Since the {\it EPOXI} CN lightcurves were undersampled and our R-band lightcurves showed much less variation than our CN lightcurves, we chose to compare our CN lightcurves with the MRI clear filter lightcurve (Figure~\ref{fig:epoxi_lc}). Both ground-based and {\it EPOXI} data are plotted with a circular aperture of radius 6 arcsec. While smaller than the aperture we employed for earlier studies, this aperture size was the best match to the ensemble of {\it EPOXI} data, where the spacecraft-centric distance varied from 0.24 AU on October 1 ($\sim$30\% farther than the geocentric distance at that time) to $<$1000 km on November 4. Times were corrected for the light travel time and fluxes were multiplied by the observer-centric distance ($\Delta$). The flux correction accounts for the changing observer-comet distance (a factor of ${\Delta}^2$) and a conversion from a fixed aperture in angular size to a fixed aperture in linear size (${\Delta}^{-1}$, under the assumption that the coma is in free expansion so the surface brightness falls off as the inverse of the distance to the comet, ${\rho}^{-1}$, at these scales). Due to differences in the bandpasses, it was necessary to scale up the {\it EPOXI} data by a factor of 6 to improve their visibility, but this had no effect on the timing of features in the lightcurve.

%FIGURE 5: KPNO/Lowell data with {\it EPOXI} lightcurve

The ground-based and {\it EPOXI} lightcurves behave similarly, with segments acquired at the same times exhibiting the same general trends, and gaps in the {\it EPOXI} dataset being filled well by the ground-based dataset. While the ground-based coverage is considerably sparser than the spacecraft coverage, the extrema appear to line up well in October. There are small systematic shifts between the datasets in November, with the ground-based lightcurve extrema occuring slightly later in November due to the significantly larger aperture size at the comet (cf. \citealt{li12}). The November extrema are aligned when we plot apertures of the same size at the comet; we do not show this because the {\it EPOXI}-comet distance changed too rapidly during this interval and the plot becomes difficult to read. Since the {\it EPOXI} spacecraft viewed Hartley 2 from a different direction than the Earth, the agreement of the shape of the lightcurve and the timing of the extrema suggest that the lightcurve shape is due to changes in the production rate. If, instead, the lightcurve shape was due to changing viewing geometry of structures in the coma, we would expect an offset in the timing of the lightcurve extrema or possibly different trends.

\subsection{Phasing of the {\it EPOXI} Lightcurve}
\label{sec:epoxi_lc_phased}
\begin{figure}[t]
  \centering
  \includegraphics[width=87mm]{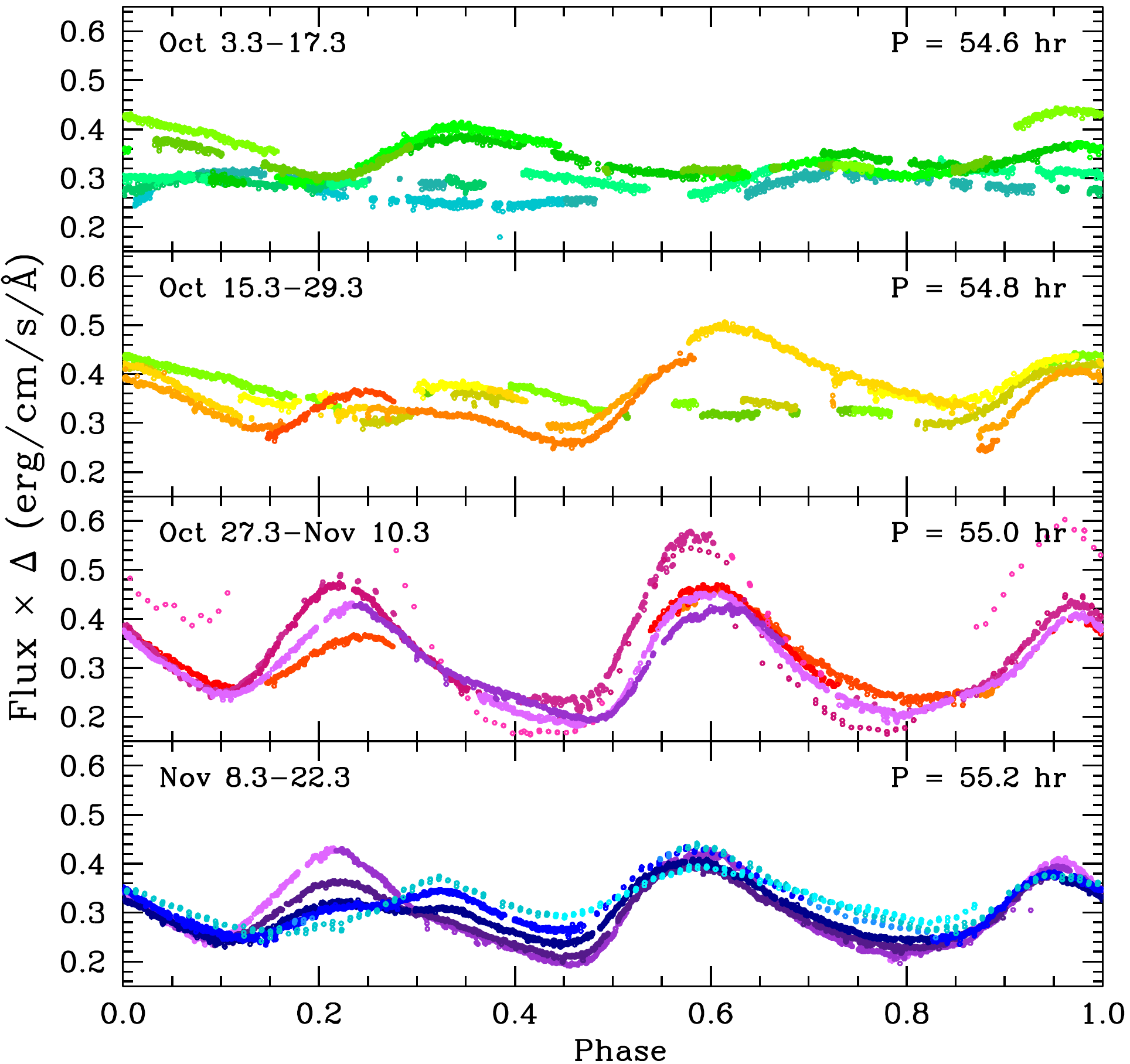}
  \caption[Phased {\it EPOXI} lightcurves]{\footnotesize{Subsections of the {\it EPOXI} clear filter lightcurve \citep{williams13} phased to an apparent triple-peaked period. The data are identical to those in Figure~\ref{fig:epoxi_lc} but are now plotted with one color per two days. The colors approximately follow a color wheel -- green, yellow, orange, red, purple, blue -- from October 3 to November 22. The date range is given in the top left of each panel, and the period to which the data are phased is given in the top right. Zero phase is set to perihelion (October 28.257). Each panel plots exactly two weeks of data with two days of overlap at the beginning/end with the panel above/below it. The dates of each panel in days relative to perihelion ($\Delta$T) are $-$25.0 to $-$11.0 (top panel), $-$13.0 to $+$1.0 (second panel), $-$1.0 to $+$13.0 (third panel), and $+$11.0 to $+$25.0 (bottom panel). Note that the data become sparse and the amplitude increases during the {\it EPOXI} close approach (November 4.587; $\Delta$T = $+$7.3).}}
  \label{fig:epoxi_phased}
\end{figure}

Having shown that our ground-based lightcurves match the {\it EPOXI} lightcurve, we now briefly consider the {\it EPOXI} lightcurve by itself. We do not intend for this to be a thorough analysis of the {\it EPOXI} data, but a means for better understanding our more sparsely sampled ground-based observations. Examination of the {\it EPOXI} lightcurve reveals that the shape and amplitude varies from one 17--19 hr cycle (the ``single-peaked'' case) to the next, but that the ensemble of features in the lightcurve repeat reasonably well over approximately 54--55 hr (the ``triple-peaked'' case). The entire {\it EPOXI} dataset cannot be satisfactorily phased with a single period, but subsets of the data can be phased reasonably well and reveal a lengthening apparent period: 54.6$\pm$0.3 hr from October 3--17, 54.8$\pm$0.5 hr from October 15--29, 55.0$\pm$0.1 hr from October 27--November 10, and 55.2$\pm$0.2 hr from November 8--22. This is shown in Figure~\ref{fig:epoxi_phased}, where we have normalized the {\it EPOXI} data as in Figure~\ref{fig:epoxi_lc} (including the scaling in order to keep the axes the same on both plots), then phased four segments to their best by-eye fits with the periods shown on the figure and zero phase set to perihelion. We have colored the points in two day steps that approximately follow the color wheel, with green in early October followed by yellow, orange, red, purple, then blue in late November; the same colors are used for the same dates in all panels.

%FIGURE 6: EPOXI lightcurve phased

The striking feature of this phased plot is that the triple-peaked lightcurve shape does not remain constant even during segments that can be phased reasonably well. While some features repeat at similar phase and amplitude, others evolve steadily from one triple-peaked cycle to the next. Note that this evolution is not due to a slight mismatch in the period used for phasing, since it occurs in only one area of the lightcurve while other areas continue to align well. This can be seen best in the bottom panel of Figure~\ref{fig:epoxi_phased}, showing November 8--22 (days from perihelion, $\Delta$T, = $+$11 to $+$25). While the maxima at $\sim$0.59 and $\sim$0.96 phase remain relatively constant in their shape, height, and location in phase, the other maximum changes steadily. The earliest data, in light pink/lavender, have the highest peak and reach it at a phase of $\sim$0.22. On the subsequent triple-peaked cycle (purple), the peak still occurs at a phase of $\sim$0.22, but falls dramatically without a significant accompanying change in the height or location of the neighboring minima. On the next two cycles (dark blue then royal blue/light blue), the maximum at $\sim$0.22 phase continues to fall and is accompanied by a second maximum at $\sim$0.33 phase that increases in height each cycle. The final cycle plotted (turquoise) shows only the peak at $\sim$0.33 phase with no evidence of the former peak at $\sim$0.22 phase. There are hints that the opposite transition between the peaks at $\sim$0.33 and $\sim$0.22 phase may be occuring at the end of the second panel and the beginning of the third panel (shades of orange), but gaps in the data prevent us from drawing firm conclusions about this.

The third panel, showing October 27 to November 10 ($\Delta$T = $-$1 to $+$13) shows a relatively small amount of variation in the lightcurve shape. There is clear evidence that the largest measured amplitudes coincide with the {\it EPOXI} encounter (November 4.587), and progressively lessen in the days before and after. The only significant variations in the shape are thought to be directly associated with this.

The first two panels, showing October 3--17 ($\Delta$T = $-$25 to $-$11) and October 15--29 ($\Delta$T = $-$13 to $+$1), have sparser coverage than the third and fourth panels, but show much more rapid evolution of features in the lightcurve. This is most evident at $\sim$0.35 phase in the top panel and $\sim$0.62 phase in the second panel, where strong peaks are seen just a few triple-peaked cycles after the lightcurve was nearly flat at the same phase. The relative lack of coherence in the phased lightcurves in these two panels is not due to poor choice of the apparent period when phasing; these were our optimal apparent periods and we could find no apparent period that resulted in features aligning as well as they do in the third or fourth panels.

While not shown, we also examined the uncalibrated {\it EPOXI} data from September 5--25 provided in the supplemental online material of \citet{belton13a}. These data were not archived with the PDS because ``a small light leak at large solar elongations allowed sunlight to enter the instrument, causing the comet to appear anomalously bright in all filters'' \citep{williams13}. In addition, the data were collected much less frequently than in October and November. We conducted similar analyses to those just discussed for the PDS data, finding a shorter apparent period, 54.0$\pm$0.5 hr from September 13--25 (the early September data could not be phased conclusively), and more rapid evolution of the triple-peaked lightcurve. The much lower sampling rate coupled with the light leak yielded poorer quality results and we therefore elected to exclude the September data from further discussion.

%\newpage
\section{DISCUSSION}
\label{sec:disc}
\begin{figure*}[ht]
  \centering
  \includegraphics[width=155mm]{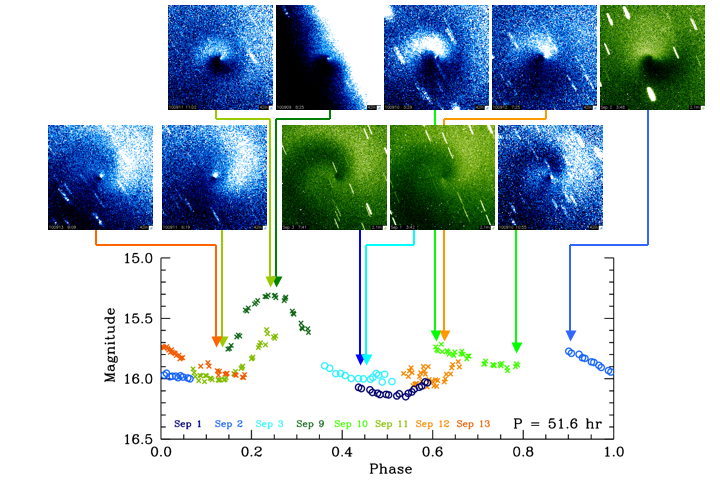}
  \includegraphics[width=155mm]{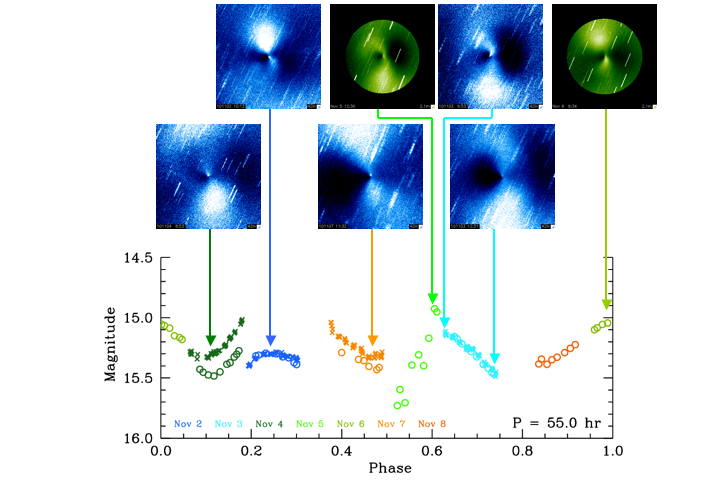}
  \caption[Triple-peaked lightcurve and coma morphology]{\footnotesize{Triple-peaked CN lightcurve and CN coma morphology for September 1--3 and 9--13 (top) and November 2--8 (bottom). The lightcurve details are given in Figure~\ref{fig:phased_lc}, and image details are given in Figure~\ref{fig:cn_pairs}. The color schemes are blue/white for Lowell and green/white for KPNO, with white representing bright regions and blue/green representing dark regions. The lightcurves were phased to the approximate best apparent triple-peaked period at the midpoint of our data. The phase for each image is noted by an arrow color coded to match that night's lightcurve. In both plots the images occurring near lightcurve maxima are shown on the top row and images occurring near the lightcurve minima are shown on the bottom row. This reveals that the morphology was much more constant through the entire apparent triple-peaked cycle in September than in November. An extremely bright star contaminated all images on September 9, making the western half of the image appear white and suppressing the coma morphology (although hints of the arc feature can still be seen). Also note that September 12 and the second half of September 13 were cloudy; as discussed in the text we have attempted to salvage these lightcurve data with comparison stars, but the coma morphology suggests that the lightcurves on these nights are unreliable.}}
  \label{fig:lc_morph}
  \label{lastfig}
\end{figure*}

Having introduced Hartley 2's coma morphology and lightcurve separately, we now consider them in combination. An obvious question is whether the apparent periods determined in each manner are the same, e.g., was similar morphology exhibited at similar lightcurve phase? We looked at the coma morphology during all maxima and minima in our data, since these are the aspects of the lightcurve that can be reliably determined each cycle despite the lightcurve's evolution. Within the level of uncertainties ($\lesssim$0.5 hr), the same features in the coma were seen at the same rotational phase throughout the subsets of data that exhibited similar overall coma morphology: August through mid-September, early- to mid-October, early November, and mid-December. The coma's appearance as a function of position in the lightcurve mirrors the the overall morphological evolution. From August through mid-September the coma looked fairly similar at the corresponding lightcurve extrema during all three single-peaked cycles within a triple-peaked cycle. In later months, the coma differed from cycle to cycle and looked most similar when integer multiples of three cycles apart. The CN lightcurves overlaid with coma morphology for early September and early November are shown in Figure~\ref{fig:lc_morph}.

%FIGURE 7: Lightcurve plus morphology

The correlation between lightcurve phase and morphology implies that both techniques should yield comparable apparent periods and that differences are likely due to the limitations of the data rather than systematic differences due to the method of measurement. This correspondence was not necessarily expected, since projection effects could cause coma morphology to repeat earlier or later on subsequent cycles. As can be seen in Figure~\ref{fig:rot_period}, previous results from smaller subsets of data generally yielded systematically longer apparent periods from morphology but, as discussed below, these can now be interpreted as due to biases in the limited ground-based sampling. Note, however, that this conclusion is not relevant for radar observations (e.g., \citealt{harmon11}) as they detected the rotation of the nucleus itself rather than variations in the activity or morphology.

One benefit of plotting the {\it EPOXI} data phased to a triple-peaked solution is that it revealed that consecutive single-peaked cycles were not identical and that some portions of the lightcurve evolved over just a few days. This was not obvious when plotting single-peaked lightcurves from the ground and explains the sometimes incompatible portions of the lightcurve we found when phasing our more sparsely sampled data. For example, the second CN panel in Figure~\ref{fig:phased_lc} shows phased data from early- and mid-September with a strong peak near 0.9 phase on September 9 and 11 (one triple-peaked cycle apart), but no obvious peak at that phase on September 12 (one single-peaked cycle later than September 11) or September 2 (10 single-peaked cycles, e.g., one single-peaked cycle plus three triple-peaked cycles, earlier than September 9). Since the coma morphology was similar in all cycles, we were able to determine a robust apparent morphological period, but the apparent period determined from the lightcurve had much larger uncertainty.

Phasing the {\it EPOXI} lightcurve to a triple-peaked period also revealed unequal spacing between lightcurve peaks. This may contribute to the large scatter in published periods since derived single-peaked periods can vary significantly depending on which lightcurve peaks are used as well as the timing of observations. As an example, using the earliest data in the bottom panel of Figure~\ref{fig:epoxi_phased} (light pink), peaks are seen at $\sim$0.22, $\sim$0.59, and $\sim$0.96 phase. This yields single-peaked periods of 20.4 hr, 20.4 hr, and 14.4 hr, all far from the combined mean single-peaked period of 18.4 hr. Just such a misleading period would be determined if our October 16 and 17 lightcurves (shown in the third row of the left column of Figure~\ref{fig:nightly_lc}) were considered in isolation since the time between troughs yields a period of $\sim$21.5 hr (note that the coma morphology did not suggest a repetition due to the variations in morphology between single-peaked cycles). In comparison, \citet{knight11b} determined an apparent period of 18.15$\pm$0.15 hr around this time based on coma morphology over several nights, while \citet{belton13a} determined an apparent period of 18.02$\pm$0.06 hr from the {\it EPOXI} lightcurve.

Shifts in the location in phase space of a lightcurve feature over a short interval, as seen in the {\it EPOXI} lightcurve when phased to the triple-peaked solution, can also distort the inferred single-peaked period. Returning to the bottom panel of Figure~\ref{fig:epoxi_phased}, periods determined using the first peak in purple (at $\sim$0.22 phase) and royal blue (at $\sim$0.33 phase) just two triple-peaked cycles apart would imply a single-peaked period of 19.4 hr for the six intervening cycles, significantly longer than the combined mean single-peaked period of 18.4 hr. The {\it EPOXI} lightcurve reveals a number of such transitions, and others undoubtedly occurred during the gaps in {\it EPOXI} coverage and before/after {\it EPOXI} observations.

In the two preceding cases, averaging over a large number of intervening cycles will reduce the scatter in the inferred single-peaked period. This explains the relatively smooth progression of single-peaked periods across the apparition that we determined from coma morphology on consecutive observing runs since these were separated by tens of intervening single-peaked cycles. However, single-peaked periods measured from just a few days of data can be heavily biased. This may explain much of the scatter in published solutions, since most were derived from data taken over just a few days.

A related uncertainty in the inferred single-peaked period comes from imperfect sampling from the ground due to the diurnal cycle. We typically observed the comet for 5--9 hr per night over 3--5 consecutive nights of a single observing run. Due simply to the cadence of our observations and the separation between observing runs, we often got little to no data on one of the three cycles that make up the triple-peaked period. We examined this by calling the three single-peaked cycles that make up one three-peaked cycle A, B, and C, arbitrarily starting A at 0:00 UT on the first night of an observing run, and defining it as extending for one single-peaked period (using the period we previously determined for that epoch), at which point cycle B started, followed by C, then by A, etc.\footnote{These cycle start/stop times are set by the dates of our observations and were not the same as those used by \citet{drahus11} to partition their data.} While we had decent sampling within all three cycles for our first two pairs of observing runs (mid-August to early September and early September to mid-September), after mid-September cycle C was almost never imaged. As just discussed, this effect is mitigated by the large number of intervening cycles, but still might yield inferred periods biased by $\sim$0.1--0.3 hr in widely spaced data. It is difficult to quantify how this affects our paired December images since {\it EPOXI} was no longer observing so we do not have full coverage of the triple-peaked lightcurve, but given that our pairs were separated by five or fewer single-peaked cycles, the uncertainty could be 1 hr or more. Note that the uncertainties listed in Table~\ref{t:period_meas} are the standard deviation of our measured periods and do not account for such biases, nor do the range of viable apparent periods given in Table~\ref{t:period_phased}.

We tested if the unequal spacing between lightcurve peaks observed in {\it EPOXI} data from October and November existed earlier in the apparition as follows. We took the interval between a matched pair of images and subtracted the apparent period determined for that epoch times ($N-$1) where $N$ is the number of intervening periods we determined for that image pair in Section~\ref{sec:morphology}. If the length of the single-peaked cycles was uniform, this ``remainder'' should be approximately equal to the apparent period; if unequal the remainder will be longer or shorter depending on the starting and ending single-peaked cycles. We found remainders that differed from the apparent period by increasing amounts during the apparition, from about 1 hr in the mid-August to early September image pairs to as much as 7 hr in  mid-October to early November image pairs. The large range of remainders should not be interpreted as being due to large uncertainties in the matching of coma morphology, as we see clustering in the remainders having the same starting and ending single-peaked cycles. That is, remainders determined from an initial image in cycle A (as defined in the previous paragraph) and a final image in cycle B were similar, remainders determined from an initial image in cycle C and a final image in cycle A were similar to each other but were different from the A-B pair remainders, etc. While all of the biases discussed above may skew the specific remainder values, the increasing range of remainders strongly suggests that the length of the single-peaked cycles were more equal in August and early September than in October and November. This finding extends earlier than the {\it EPOXI} monitoring and therefore is a valuable constraint for future efforts to understand the component periods of NPA rotation (which we presume are responsible for the observed behavior).

Despite the large uncertainties in our periods after early November, our combined data suggest the apparent period changed little after the {\it EPOXI} flyby. This is consistent with \citet{belton13a}'s analysis of the {\it EPOXI} spacecraft data which found a period of 18.63$\pm$0.02 hr on November 11 and 18.61$\pm$0.09 hr on November 20. In light of our better understanding of how sampling bias can skew period determinations, it is also probably consistent with our earlier estimate of an apparent period ``near 19 hr'' on December 9--10 \citep{knight11b} and \citet{cbet2589}'s report that the apparent period increased from 18.2 to 19 hr from October 29 to December 7 (the 0.3 hr error bar on their period from the first two weeks of data, when the comet was brighter, implies their uncertainties by early December were at least this large).

The gas production rates of Hartley 2 peaked 0--20 days post-perihelion (e.g., \citealt{knight13}). Assuming a simple model where the outgassing torque is proportional to the gas production rate, \citet{knight11b} argued for non-linear changes in the apparent period, and a decrease in the rate of change of the period was predicted starting from early November (see their Figure 6). However, the extent of the flattening of the rate of change of the period appears to be much more than suggested there. The rate of change of the rotational state, including those of the component periods, depends on the net component torques acting on the nucleus due to outgassing. Torques can be due to outgassing from specific source regions on the surface or due to much more widely distributed activity on the surface of the nucleus. Regardless of the primary cause, the outgassing pattern and the resultant net torques can easily change due to different insolation patterns as a comet moves along in its orbit. In other words, the net component torques acting on the nucleus are not necessarily proportional to the gas production rate or to the level of surface activity (cf. \citealt{samarasinha13}). Due to this complexity, one should be cautious in directly correlating this flattening observed for the period in November and December to the activity of the nucleus. The apparent disconnection between the torquing and the production rates is another constraint for future modeling efforts.

We found a steady (within the uncertainties) increase in the apparent triple-peaked period when phasing the {\it EPOXI} data, in contrast to earlier analyses by \citet{ahearn11a} and \citet{belton13a} who found more variability in the period during the same interval. This is likely due to how each analysis technique handles the most rapidly evolving segments of the lightcurve. When visually assessing goodness of fit, we were cognizant of the evolution of the lightcurve from cycle to cycle. Thus, we disregarded regions in obvious transition, instead focusing on those regions that remained relatively unchanged during a given interval. Since the lightcurve was continuously evolving, focusing on alignment of features in the more rapidly evolving regions would have yielded much larger uncertainty and, likely, a more chaotic trend in the apparent period during the {\it EPOXI} observing window. In contrast, the power spectra technique employed by \citet{ahearn11a} and \citet{belton13a} treats all data equally. \citet{belton13a} found very small uncertainties ($<$0.07 hr) when the lightcurve was most stable, in late October and early November, but found apparent periods with more variation and considerably larger uncertainties (0.86--5.76 hr) in September and early October, when evolution of the lightcurve was significant, the data were sampled less frequently, and (in September) there was a light leak that varied over time. The large uncertainty in September makes it difficult to determine a clear trend in this component period in either \citet{ahearn11a}'s or \citet{belton13a}'s analyses. We surmise that Hartley 2's component periods changed steadily during the apparition and therefore conclude that the apparent triple-peaked periods we determined from visually phasing the data are more plausible.

The fact that the triple-peaked and single-peaked apparent periods are almost exactly in a 3:1 ratio throughout the {\it EPOXI} observing window should not be surprising -- the triple-peaked apparent period has three peaks and by definition should yield a single-peaked apparent period that is 1/3 as long. What is surprising is that two of the underlying component periods of NPA rotation appear to have been close enough to a 3:1 ratio (or some alias such as 3:2) that the triple-peaked lightcurve exhibited minimal evolution during the third panel of Figure~\ref{fig:epoxi_phased} (October 27 to November 10). The lightcurve clearly showed more rapid evolution in the surrounding weeks (shown in the first, third, and fourth panels of Figure~\ref{fig:epoxi_phased}). This seems to imply that at least one of the component periods of NPA rotation was changing such that their relative ratio was approximately an integer multiple sometime between October 27 and November 10, and was further away from an integer multiple both earlier and later. A similar conclusion is reached using the ratio of \citet{belton13a}'s ${\nu}_4$ and ${\nu}_1$, which were in a 3:1 ratio approximately October 27. The transition in lightcurve shape is a hallmark of NPA rotation, and similar behavior was seen in the lightcurve of 1P/Halley (\citealt{schleicher90}; Schleicher et al., in prep.). This serves to remind us that the apparent periods we are measuring are not necessarily the component periods of the NPA rotation, but the interaction of the underlying periods manifesting themselves as repetition in coma morphology and brightness.

An additional reminder of the evolving ratio of the component periods is the variation in morphology at the the same lightcurve extrema seen in Figure~\ref{fig:lc_morph}. In September, the morphology near all three peaks in the lightcurve is similar, as is the morphology near all three troughs. In contrast, the morphology varies from one peak to the next in November and the bulk brightness flips hemispheres during the middle cycle. While beyond the scope of the current paper, this behavior will be an important constraint for future modeling efforts.

\section{SUMMARY}
\label{sec:summary}

In this work we have combined observations of Hartley 2 during the 2010 apparition obtained at Kitt Peak National Observatory and Lowell Observatory. The two datasets were not contemporaneous except at the time of the {\it EPOXI} flyby, and the combined dataset roughly halves the time between consecutive observing runs as compared to data collected at a single observatory. This allowed determinations of intermediate periods that were not possible between consecutive runs with the same telescope. We measured the apparent period of CN coma morphology repeatability between runs, yielding apparent periods additional to those previously reported from our individual datasets \citep{knight11b,samarasinha11}. We also measured inner coma lightcurves in both continuum (R-band) and gas (CN) within and between observing runs, the first such lightcurves of Hartley 2 obtained from the ground to be published. Phasing of these lightcurves yielded additional apparent periods and revealed that the coma morphology and lightcurves were in sync.

Our new periods are consistent with the roughly linear increase in apparent period reported by previous authors up to mid-November. The post-{\it EPOXI} encounter periods suggest that the period flattened out or possibly even decreased slightly in late November and December. The next favorable observing window for Hartley 2 will occur in mid-2016, prior to the onset of strong activity. Observations at this time should reveal the current period and test how it has changed since late 2010. However, as this will still be at relatively large heliocentric distances, the brightness may be dominated by the nucleus, in which case large aperture telescopes will be required and the lightcurve may display different characteristics than were seen in the activity dominated lightcurve of 2010. Unfortunately, Hartley 2 next reaches perihelion on the far side of the Sun in April 2017, so Earth-based observations designed to look for further changes in the period will likely not be possible until early 2018 and will again require large aperture telescopes.

We also examined the {\it EPOXI} lightcurve in order to better understand our own data. Phasing of the {\it EPOXI} data to a triple-peaked period revealed that consecutive single-peaked cycles had different lightcurve shapes and separations between peaks that approximately repeated every third cycle. Segments of the triple-peaked lightcurve evolved from one triple-peaked cycle to the next while other segments remained approximately constant for many cycles. These behaviors can explain much of the scatter between apparent periods measured at similar epochs in our own and other datasets. We showed that limitations of undersampled datasets can lead to differences from the true period of 1 hr or more, making it more important to have longer baselines.

Despite Hartley 2 having been extensively studied from the ground and space in support of {\it EPOXI}, its rotation is not yet fully understood. The evolution of the lightcurve shape and coma morphology seem to indicate that the relative ratios of the component periods of NPA rotation changed during the apparition. However, specific change(s) corresponding to component periods are currently not known. At present there is also no consensus rotational angular momentum vector (RAMV), with multiple authors suggesting very different RAMVs (e.g., \citealt{ahearn11a,harmon11,waniak12,knight13,belton13a}). Furthermore, the type of NPA rotation (LAM versus SAM) is not yet conclusively known. Hartley 2's NPA rotation is challenging to model due to the large number of free parameters, but numerical simulations may yet arrive at a solution that matches the coma morphology and replicates the complex lightcurve. Such a result would then be very useful for planning a future spacecraft mission to this compelling comet.

\section*{ACKNOWLEDGMENTS}
We thank the anonymous referee for useful sggestions that improved the manuscript. We thank Drs. Michael A'Hearn, Tony Farnham, and Alan Gersch for being part of the  Kitt Peak observing team and Edward Schwieterman for being part of the Lowell observing team. We also thank Dr. Farnham for carrying out the absolute calibration for the Kitt Peak narrowband HB filter images. We thank Michaela Fendrock for help with lightcurve analysis.

M.M.K. is grateful for office space provided by the University of Maryland Department of Astronomy and Johns Hopkins University Applied Physics Laboratory while working on this project. D.G.S. and M.M.K. were supported by NASA's Planetary Astronomy Program grant NNX11AD95G and NASA Planetary Atmospheres grant NNX14AH32G. B.E.A.M. and N.H.S. were supported by NASA Planetary Atmospheres grant NNX12AG56G and NASA Planetary Astronomy grant NNX09AB42G.

This is PSI contribution TBD.

\end{document}